\documentclass[twocolumn,aps,pra,superscriptaddress,showpacs,showkeys]{revtex4}
\usepackage{textcomp}
\usepackage{latexsym}
\usepackage{graphicx}
\usepackage{units} 

\begin{document}
\title{Atmospheric Channel Characteristics for Quantum Communication\\with Continuous Polarization Variables}
\author{Bettina Heim }\email{bettina.heim@mpl.mpg.de}\author{Dominique Elser}\author{Tim Bartley}\author{Metin Sabuncu}\author{Christoffer Wittmann}\author{Denis Sych}\author{Christoph Marquardt}\author{Gerd Leuchs}

\affiliation{Institute of Optics, Information and Photonics, University of Erlangen-Nuremberg\\ Staudtstr.~7/B2, 91058 Erlangen }
\affiliation{Max Planck Institute for the Science of Light\\G\"unther-Scharowsky-Str.~1, Bau 24, 91058 Erlangen}


\begin{abstract}
We investigate the properties of an atmospheric channel for free space quantum communication with continuous polarization variables. In our prepare-and-measure setup, coherent polarization states are transmitted through an atmospheric quantum channel of \unit[100]{m} length on the roof of our institute's building. The signal states are measured by homodyne detection with the help of a local oscillator (LO) which propagates in the same spatial mode as the signal, orthogonally polarized to it. Thus the interference of signal and LO is excellent and atmospheric fluctuations are auto-compensated. The LO also acts as a spatial and spectral filter, which allows for unrestrained daylight operation. Important characteristics for our system are atmospheric channel influences that could cause polarization, intensity and position excess noise. Therefore we study these influences in detail. Our results indicate that the channel is suitable for our quantum communication system in most weather conditions.
\end{abstract}

\maketitle
\section{Introduction}
\label{intro}
Quantum communication describes the distribution of quantum states between two parties, traditionally named Alice and Bob. These states can for example be entangled~\cite{einstein35} states, providing the basis for various protocols such as quantum teleportation~\cite{bennett93} or quantum dense coding~\cite{bennett92}. Many of the initial research projects used discrete quantum variables. Later also continuous variables have proven suitable for quantum communication (for a review see~\cite{andersen09}).

Quantum key distribution (QKD)~\cite{gisin02,scarani08} is a further important branch of quantum communication and concerns the establishment of a secret key jointly between Alice and Bob with the help of a quantum channel. The security is based on the laws of quantum mechanics. In principle unconditional security can be achieved. Any two non-orthogonal quantum states suffice to ensure secure key distribution~\cite{bennett92x}. This holds as long as the detection matches the quantum state emitted by the source. A single photon detector e.g.~matches weak coherent states as long as the probability for multiphoton events is low enough. For higher multiphoton probabilities~\cite{grosshans03} or even bright polarization states~\cite{lorenz04,elser09} a single photon detector can not be used. In such scenarios, however, photon number resolving detectors, homodyne or heterodyne detectors are a better match, promising unconditionally secure key distribution. 

Free space QKD over an atmospheric channel was first demonstrated in 1996~\cite{jacobs96}. Since then, a number of  prepare-and-measure as well as entanglement based schemes have been implemented in free space (for a review see~\cite{scarani08}). The current world record in distance is \unit[144]{km}~\cite{schmitt-manderbach07,ursin07a} and satellite quantum communication is already in preparation~\cite{villoresi08,perdiguesarmengol08}. All of these aforementioned systems use single-photon detectors and therefore have to employ spatial, spectral and/or temporal filtering in order to reduce background light. In our system, we use an alternative approach: with the help of a bright local oscillator (LO), we perform homodyne measurements on weak coherent polarization states \cite{elser09}. We focus on the characterization of the quantum channel, which is a \unit[100]{m} free space link on the roof of our institute's building. 

In classical free space communication systems using homodyne detection (e.g.~\cite{lange06}), producing the LO locally at the receiver is appropriate. In QKD, on the other hand, the requirements for detection efficiency are more stringent, as fragile quantum states are transmitted. Thus we developed a protocol using the polarization degree of freedom to multiplex signal and LO~\cite{lorenz04}. The LO is produced by Alice and propagates in the same spatial channel mode as the signal.

In quantum mechanics, polarization is conveniently described by the quantum Stokes operators, that are the quantum counterpart of the classical Stokes parameters~\cite{Stokes1852}. The Stokes operators are introduced and defined for example in~\cite{korolkova02}. 

In a homodyne detection of the Stokes parameters, the co-propagation of signal and LO leads to an intrinsically excellent spatial interference between the two. This translates to a high detection efficiency without any additional interference stabilization. For our free space system, there are also advantageous side effects of this co-propagation: firstly, the LO acts as a spatial filter, such that only those photons, that are spatially mode-matched to it will result in a significant detector signal. Unlike in single photon experiments there is no need for spatial filtering by pinholes or fibers. Secondly, the LO facilitates spectral filtering, as the beat-note of  signal and LO, interfering at a polarizing beam splitter (PBS), can be electronically filtered at the detector. The detection bandwidth can thus be adjusted precisely and background light outside this range does not disturb the measurement. Finally, absolute phase fluctuations in the channel are auto-compensated, as they are identical for signal and LO. 

The theory for the propagation of classical light through turbulent atmosphere including diverse phenomena such as beam wander or beam spreading has been investigated in e.g.~\cite{tatarskii71,fante75,fante80}. However, effects on quantum continuous variable states have only recently been studied in this context~\cite{heersink06,dong08,semenov09}. Influences of the atmospheric channel may cause polarization and intensity excess noise under certain conditions. Both might fundamentally compromise the security of a QKD system and generally degrade transmitted nonclassical states. Atmospheric noise typically is of non-Gaussian character (e.g. on-off noise). Squeezed and entangled states that were degraded by this noise can be distilled with Gaussian operations~\cite{heersink06,dong08}. Intensity noise can easily stem from practical issues such as finite aperture size leading to fluctuations of the detected intensity. Such effects have thus to be studied and characterized in detail in order to determine if they could affect the quality of the quantum communication channel. In security analysis of QKD systems, all excess noise is considered to originate from Eve's interactions. In a worst case scenario, strong noise effects would result in the fact that no key can be established. 
\section{Experimental Setup}\label{sec:1}
\begin{figure}[t!]
\begin{center}
\includegraphics[width=0.45\textwidth]{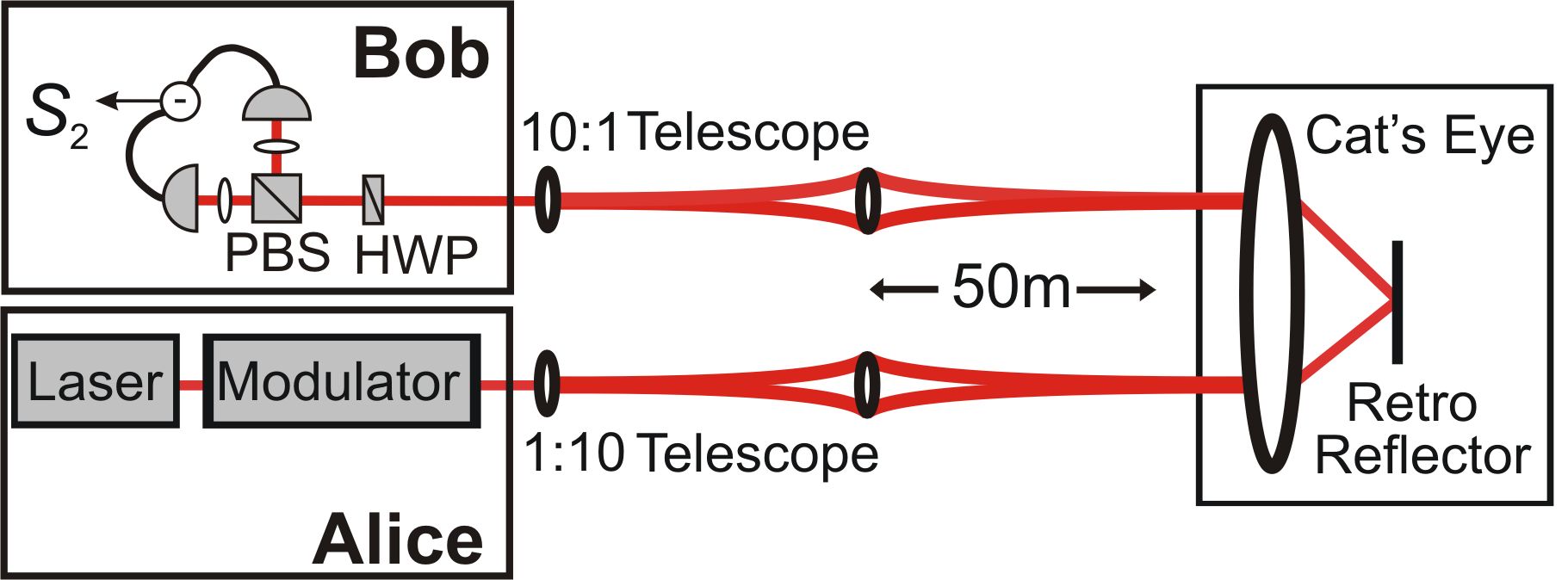}
\caption{Experimental setup for our QKD feasibility studies~\cite{elser09}: Alice's laser emits a linearly polarized CW beam which later serves as a local oscillator (LO) for Bob's measurements. In terms of Stokes operators, the local oscillator is $\hat S_1$-polarized. Alice's modulator generates a weak signal that Bob then measures by an $\hat S_2$ Stokes detection. In-between, the beam is expanded and sent to a retro reflector at a distance of \unit [50] {m}. After reflection, Bob's telescope again reduces the beam diameter. PBS: polarizing beam splitter, HWP: half wave plate.
}\label{fig:figure1}
\end{center}
\end{figure}
\subsection{Quantum state measurements}
The setup shown in figure \ref{fig:figure1} was used for our QKD-feasibility studies \cite{elser09} and follows the principles of our earlier laboratory work~\cite{lorenz04,lorenz06}. We use a grating-stabilized CW diode laser, whose wavelength of 809 nm lies within an atmospheric transmission window. A linearly polarized laser beam ($\hat S_1$ in terms of Stokes operators) is emitted by Alice and later serves as a LO in Bob's measurement. A modulator is used to generate the coherent signal states. (A magneto-optical-modulator (MOM) for example employs the Faraday effect to tilt the linear polarization by small amounts.) The weak signal component of a mean photon number of typically less than one photon per pulse is located in the same spatial mode as the LO, but is polarized orthogonally to it. After expanding the beam by a telescope, the signal/LO beam is sent over the roof of our institute's building and retro reflected after \unit [50]{m}. Bob reduces the beam diameter with a telescope and then performs a Stokes measurement of the $\hat S_2$-operator to detect the signal states.
\subsection{Setups for Different Noise Measurements}
Here we present measurements of the polarization, position and intensity excess noise properties of the atmospheric channel.
\subsubsection {Atmospheric polarization noise}
In previous work~\cite{elser09,bartley08,elser08} we investigated the polarization excess noise introduced by the channel. For an alphabet using two coherent polarization states we compared the distributions of $\hat S_2$-Stokes measurements of the signal states before and after transmission through the channel. Additional polarization noise introduced by the channel would broaden the measurement distribution. The work in~\cite{bartley08,elser08} showed, that this is not the case. Measurements of the RF frequency spectrum of unmodulated beams that were sent through the atmosphere also allow us to identify the frequency range above 10 kHz to be essentially noiseless \cite{elser09}.
\subsubsection{Atmospheric intensity noise}\label{exp_intnoise}
Atmospheric intensity noise can be measured by direct detection of the beam. For calibration, we compare the noise of a beam sent through the atmosphere with a beam sent over the optical table. The intensity noise is recorded by a spectrum analyzer. These measurements are sensitive to fluctuations of the laser's intrinsic excess noise which we monitored accurately when recording the spectra. We use low noise detectors whose electronic noise is significantly smaller than the shot noise, thus allowing us to measure at the quantum noise limit.
\begin{figure}
\begin{center}
\includegraphics[width=0.49\textwidth]{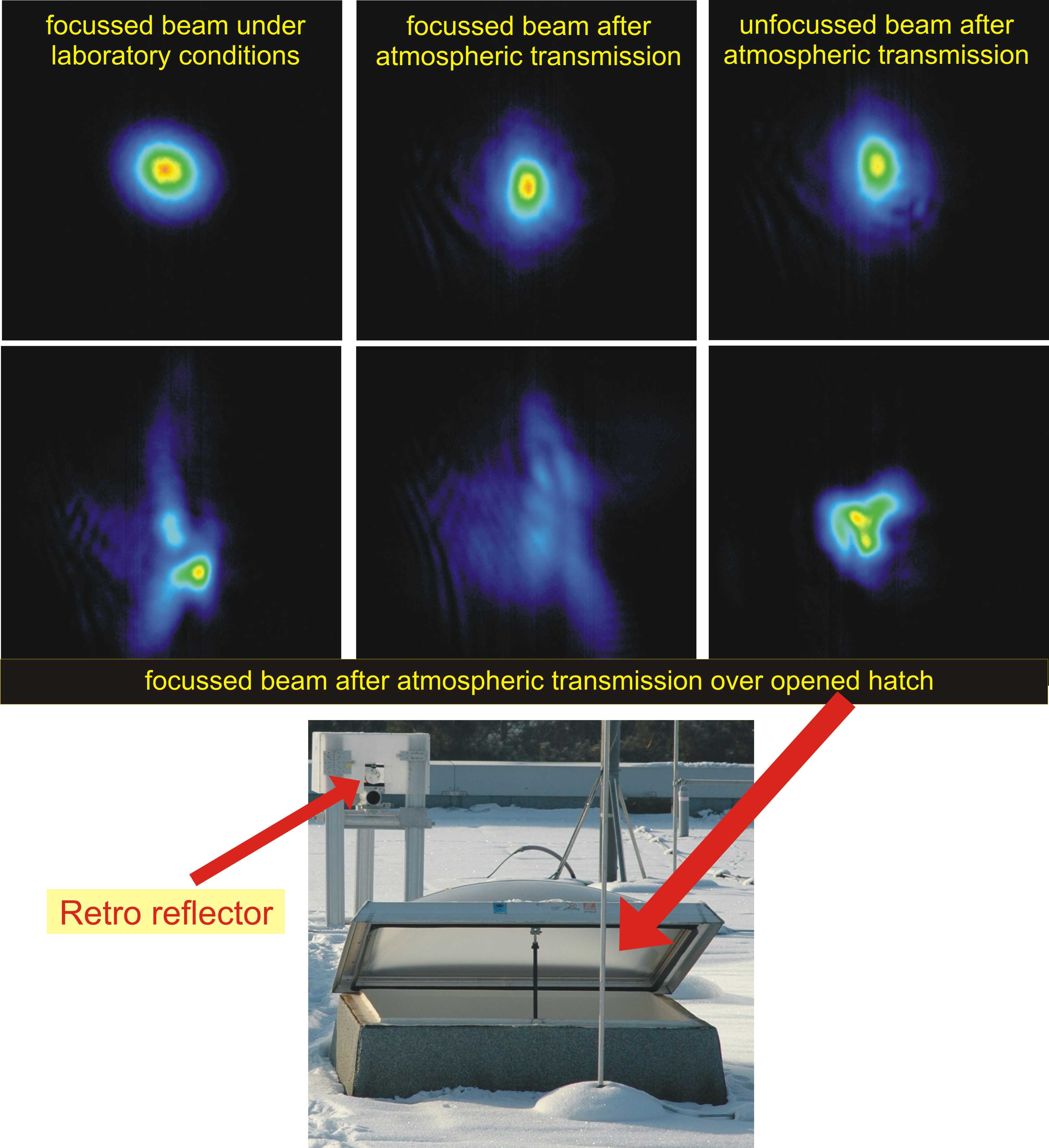}
\caption{2D-Beamprofiles under several conditions. Before being sent over the roof, the beam is in a near TEM$_{00}$ mode (upper left picture). After transmission through the channel the beam profiles are slightly distorted, but the intensity distributions along the two main beam axis are still approximately Gaussian. Strong beam distortions are caused by opening a hatch over which the beam passes on its way to and back from the retro reflector. Plots thereof, recorded at different instances of time, are shown in the second row. All beam profiles were recorded at an exposure time of \unit[20]{\textmu s}.
}\label{fig:beamprofiles}
\end{center}
\end{figure}
\subsubsection{Atmospheric beam jitter}\label{exp_posnoise}
We used a beam profiling system to compare the changes of the spatial beam profile caused by the atmospheric channel. For comparing both, the outgoing and the returning beam were detected with the help of a CCD camera (Metrolux ML3743). The pictures then were analyzed by the Metrolux BeamLux II software package. Figure \ref{fig:beamprofiles} shows some typical spatial beam profiles under different conditions. Sequences of pictures were taken at an exposure time of \unit [20]{\textmu s}. 
\section{Results and Discussion}
\label{sec:result}
\begin{figure}[t!]
\begin{center}
\includegraphics[width=0.49\textwidth]{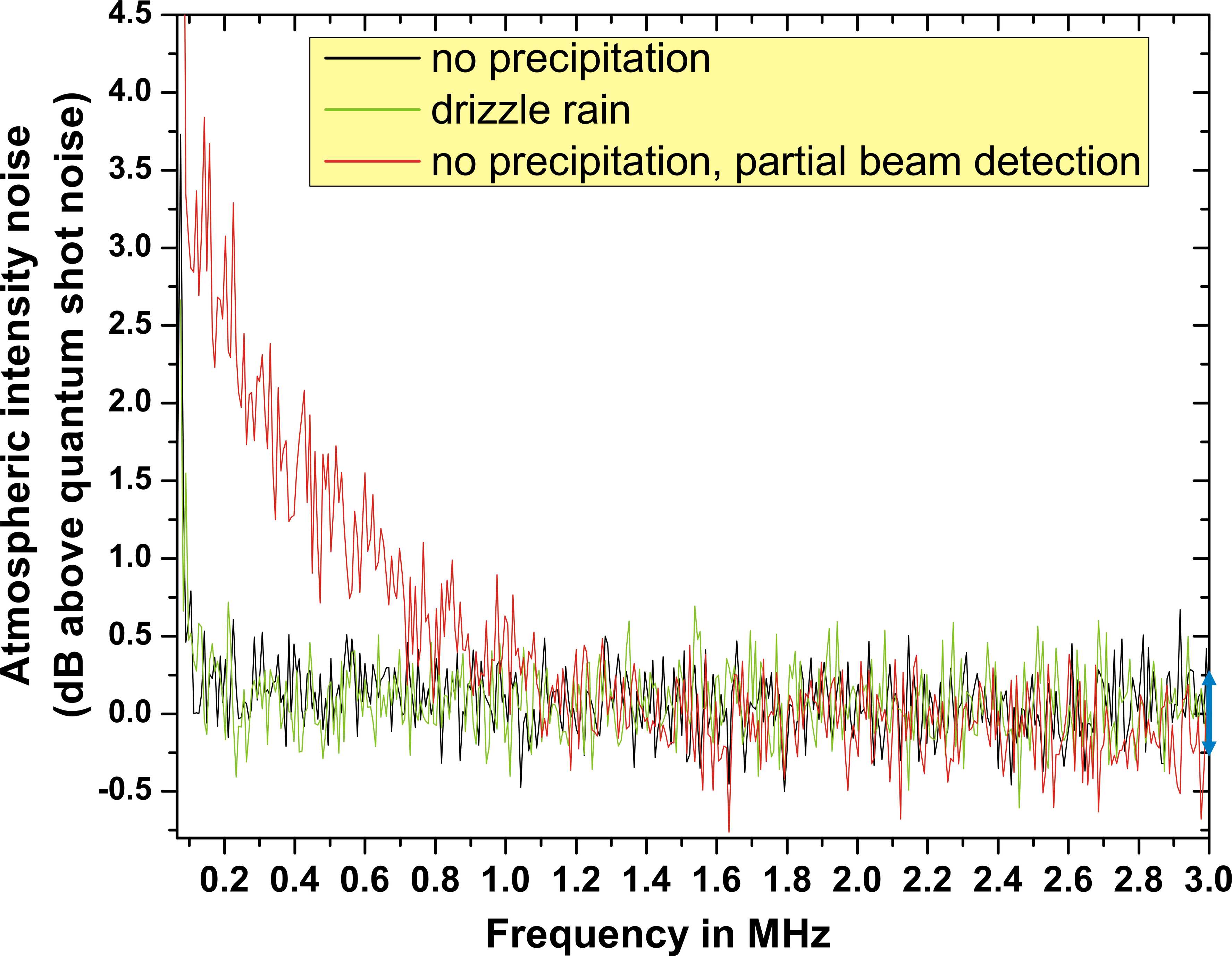}
\caption{Intensity noise measured by a direct detection. All curves arise from an averaging over several measurements and are normalized to a quantum noise limited reference beam. The resolution bandwidth for the measurements consisting of 401 points was \unit[10]{kHz}, the video bandwidth \unit[100]{Hz}. On the right hand side of the plot, the measurement accuracy of \unit[5]{\%} is shown in blue.}
\label{fig:intnoise}
\end{center}
\end{figure}
In the following we will concentrate on atmospheric intensity noise and beam jitter (atmospheric polarization noise has been investigated in~\cite{elser09,bartley08,elser08}.

\subsection{Atmospheric intensity noise}\label{res_intnoise}
Measurements of the intensity noise were performed by detecting the amplified photocurrents of one photodiode (Hamamatsu, S3399, active area \unit[7]{mm$^2$}, diameter \unit[3]{mm}), and comparing the spectrum of an unmodulated beam transmitted over the optical channel with that of a reference beam over the table. Constant attenuation was compensated for by setting the optical power in front of the photodiode to the same value of \unit[650]{\textmu W} in both cases. In figure~\ref{fig:intnoise}, it can be seen that there is no atmospheric excess noise measured for beams that are sent through the optical channel in good weather conditions (dry and sunny) as well as in light rain. This is valid for frequencies above the current QKD modulation frequency of \unit[1]{MHz}. The measurement accuracy of 5\% accomodates the fact that the beam moves on regions of the photodiode with slightly different sensitivities. Additionally, small intensity fluctuations of the laser are included. We can infer that the scattering effects of the atmosphere do not cause a measurable beam broadening or spatial beam jitter, and thus, the beam hits the photodiodes as well as it does when sent over the optical table. 

Intensity excess noise can occur if the collimated beam after Bob's telescope is detected without focussing it onto the photodiodes. Then the spatial beam jitter caused by the atmosphere exceeds the active area of the photodiodes and thus leads to partial detection noise (red curve in figure~\ref{fig:intnoise}). An estimation of this noise based on fluctuations of the beam centers will be given in the next section.

\begin{figure*} [t!]
\begin{center}
\includegraphics[width=0.9\textwidth]{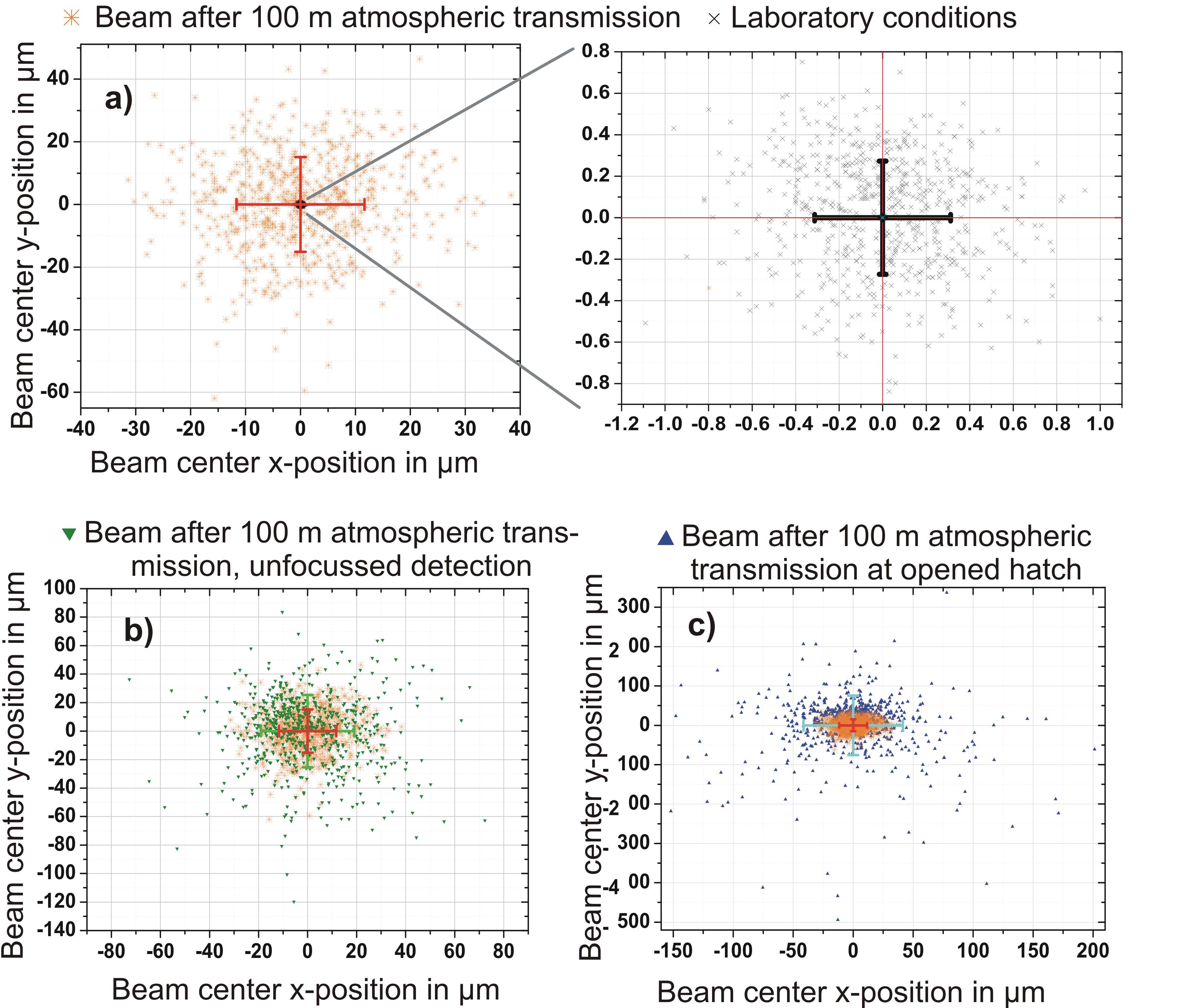}
\caption{$(x,y)$-plots of the the beam centers and standard deviations of sequences of 650 beam profiles, recorded with an exposure time of \unit[20]{\textmu s}. The mean values of the beam centers are shifted to (0,0) for each plot, the standard deviations are shown in colors corresponding to the particular beam centers. Plot a) shows the comparison of a beam that was sent over the optical table with one that passed through the atmospheric channel, both being focussed on the camera while recorded. As expected, the fluctuations are much larger through the atmosphere. In the lower part, the difference between a focussed and an unfocussed "atmospheric" beam is demonstrated (part b) ), corresponding to the intensity excess noise shown in red in figure \ref{fig:intnoise}. In part c), we compare this "atmospheric" beam to one having passed directly over a hatch, whereby the temperature gradient between inside the building and outside caused strong atmospheric fluctuations (these measurements were taken during winter time).
}\label{fig:xyplot}
\end{center}
\end{figure*}

\subsection{Atmospheric beam jitter}\label{sec:bcfluc}
Figure~\ref{fig:xyplot} show the beam centers and standard deviations of sequences of beam profiles (typical spatial intensity distributions are shown in figure \ref{fig:beamprofiles}).

Part a) in figure~\ref{fig:xyplot} shows the comparison of a beam that was sent over the optical table with one sent over the roof, both of which were focussed on the camera when recorded. As expected, the fluctuations of the beam centers are much higher for atmospheric transmission, but still small enough to be compensated by the aperture of the photodiodes. This is confirmed by the fact, that in this case no intensity excess noise is shown in figure~\ref{fig:intnoise}. 
One can estimate the relative quantum shot noise of these states by $\frac{\sqrt{\left<n\right>}}{\left<n\right>}= 2\times 10^{-7}$. The mean photon number~$\left< n\right >$ of the \unit[650]{\textmu W}-beams per measurement period is approximated by dividing the total detected energy per period  ($E_{\textnormal{total}}$) by the energy of one photon of \unit[809]{nm} ($E_{\textnormal{photon}}$):
\begin{equation}
\left<n\right>=\frac{ E_{\textnormal{total}} }{E_{\textnormal{photon}}}= \frac{P_{\textnormal{opt}} \frac{1}{\textnormal{VBW}}}{\unit[2.45\times 10^{-19}]{Ws}}=2.65 \times 10^{13}
\end{equation}
with a video bandwidth (VBW) of the RF spectrum analyzer of \unit[100]{Hz}.

By a numerical evaluation we estimate the intensity noise caused by the atmosphere for a focussed detection. The calculations are based on the measured beam center fluctuations and on values for beam diameters that we also gained from the spatial beam profiles. They refer to a certain frequency: \unit[50]{kHz}, at which the camera measurements were performed. In our calculation, we integrate over the intensity distributions of beam profiles within a region defined by the size of the photodiodes. The resulting intensity value then is normalized to that of non-cropped beams. Aligning inaccuracies of \unit[0.2]{mm} are also included in our calculations. We assume a Gaussian intensity distribution and a mean beam diameter of \unit[0.98]{mm}.

Using the calculations explained above, we compare a beam whose center is shifted by \unit[0.0134]{mm}, the mean standard deviation of the beam center fluctuations (see figure~\ref{fig:xyplot} a) ), to a centered beam. After normalization to the intensity within the size of the photodiodes, the result quotes a value for the relative intensity noise which is around $7\times 10^{-8}$. This value lies below the quantum shot noise estimated above, that marks the quantum mechanical limitation of our measurement accuracy. Thus it is too small to be detected which is in agreement with figure~\ref{fig:intnoise}.

As quoted in section~\ref{res_intnoise} and shown in figure~\ref{fig:intnoise}, intensity noise can occur by a detection of the collimated beam directly after Bob's telescope, without using a focussing lens. This beam is broadened compared to a focussed one and its beam center fluctuations are slightly higher, shown in figure~\ref{fig:xyplot} b). We perform the same evaluation as for the focussed beam above, resulting in an intensity noise of $4.4 \times 10^{-7}$ at \unit[50]{kHz}, which is about twice the estimated value for the relative quantum noise ($2\times 10^{-7}$). This is in agreement with figure~\ref{fig:intnoise}, showing the intensity noise for an unfocussed beam to be about \unit[3]{dB} higher than shotnoise for the lowest measured frequencies at around \unit[80]{kHz}. As the detection bandwidth was limited we couldn't perform the intensity noise measurements all the way down to \unit[50]{kHz}. We expect a slight further increase of the noise for smaller frequencies. Thus, the estimation is in good over all agreement with the measurements.  We want to stress, that even though this effect is small, it is still observable because of the low-noise properties of the detectors having an electronic noise level well below the quantum noise limit.

 The beam center fluctuations are even larger when a hatch (see figure~\ref{fig:beamprofiles}) is open, over which the beam passes on its way to and from the retro reflector (see figure~\ref{fig:xyplot} c)). In this particular case, atmospheric fluctuations are dramatically increased. This would cause further intensity noise when the beam is detected at the photodiode. Hence we are now working on an optimised detection system to improve free space beam capture. The use of improved optical tapers can combat strong combined spatial and angular fluctuations of the incident beam, better than a single lens could do~\cite{bartley09}. We have experimental evidence for the non-Gaussian character of the noise, which will be reported elsewhere. 
 
\section{Conclusion and Outlook}
Within the framework of the first demonstration of continuous variable quantum communication through a real atmospheric channel, we investigated different channel noise properties. We precisely characterized atmospheric intensity fluctuations by quantum-noise limited measurements. Our results indicate that in good weather conditions and with an appropriate design of sending and receiving optics, channel influences like polarization, intensity and position noise are sufficiently low to allow for quantum state transmission and QKD operation at daylight.  For our \unit [100]{m} link, there was no need for active beam stabilization, as beam jitter effects caused by the atmosphere could be compensated by appropriate design of the passive optical components. For an extended link of \unit[1.6]{km}, on which we are working currently, active stabilization is probably necessary. Monitoring the bright LO can provide us with a control signal for active beam stabilization. Additionally, to synchronize Alice's and Bob's stations, one can interrupt the LO in regular time intervals. Each time when switching on the LO marks the beginning of a new signal frame. This will fulfill the same task as the timing pulses in other free space QKD setups, e.g.~\cite{hughes02}. Furthermore, we plan to increase the pulse rate of the quantum states and implement more complex signal alphabets~\cite{sych09} in QKD.
\begin{acknowledgments}
Denis Sych acknowledges the Alexander von Humboldt Foundation for support through a fellowship.
\end{acknowledgments}

\end{document}